\documentclass[a4paper,11pt]{article}
\usepackage{pos}

\newcommand{\OpenLoops}{{\rmfamily\scshape OpenLoops}}

\newcommand{\f}[2]{\frac{#1}{#2}}

\newcommand{\ssst}[1]{\scriptscriptstyle{\text{#1}}}
\newcommand{\nosss}[1]{#1}

\newcommand{\bea}{\begin{eqnarray}}
\newcommand{\eea}{\end{eqnarray}}
\newcommand{\be}{\begin{equation}}
\newcommand{\ee}{\end{equation}}
\newcommand{\ba}{\begin{align}}
\newcommand{\ea}{\end{align}}
\newcommand{\beas}{\begin{eqnarray*}}
\newcommand{\eeas}{\end{eqnarray*}}
\newcommand{\bes}{\begin{equation*}}
\newcommand{\ees}{\end{equation*}}
\newcommand{\bas}{\begin{align*}}
\newcommand{\eas}{\end{align*}}

\newcommand{\eps}{{\varepsilon}}

\newcommand{\lb}{\left(}
\newcommand{\rb}{\right)}

\newcommand{\momk}[1]{k_{\nosss{#1}}}

\newcommand{\mass}[1]{m_{\nosss{#1}}}
\newcommand{\heli}{h}

\newcommand{\helicheck}{\check h}

\newcommand{\helihat}{\hat h}

\newcommand{\momq}{\bar{q}}

\newcommand{\calM}{\mathcal{M}}
\newcommand{\calMCT}{\mathcal{M}^{\ssst{(CT)}}}
\newcommand{\calN}{\mathcal{N}}

\newcommand{\barM}{\bar{\mathcal{M}}}

\newcommand{\calV}{\mathcal{V}}
\newcommand{\calW}{\mathcal{W}}
\newcommand{\seg}{S}

\newcommand{\calU}{\mathcal{U}}
\newcommand{\helicheckloc}{\tilde h}

\newcommand{\segment}[2]{\seg_{#2}^{(#1)}}

\newcommand{\helisegment}[2]{\heli_{#2}^{(#1)}}

\newcommand{\helic}[1]{\heli^{(#1)}}                   
\newcommand{\helig}{\heli}                             
\newcommand{\helipc}[2]{\helihat_{#2}^{(#1)}}          
\newcommand{\helipcc}[2]{\helicheck_{#2}^{(#1)}}       
\newcommand{\helipccloc}[2]{\helicheckloc_{#2}^{(#1)}} 

\newcommand{\numc}[1]{\calN^{(#1)}}                 
\newcommand{\numpc}[2]{\calN^{(#1)}_{#2}}           
\newcommand{\numpi}[1]{\calU^{(1)}_{#1}}           
\newcommand{\numpd}[1]{\calU^{(123)}_{#1}}           
\newcommand{\numinter}[1]{{\calU_{#1}^{(13)}}}                

\newcommand{\fullamp}[2]{{\calM}_{{#1},{#2}}}

\newcommand{\colfac}[2]{{C}_{{#1},{#2}}}
\newcommand{\vertex}[1]{\calV_{#1}}
\newcommand{\denc}[1]{\mathcal{D}^{(#1)}(\bar q_{#1})} 
\newcommand{\col}{\mathrm{col}}

\newcommand{\re}{\mathrm{Re}}

\newcommand{\rd}{\mathrm d}

\definecolor{bluemar}{rgb}{0,0,.5}
\definecolor{redmar}{rgb}{.8,0,0}
\definecolor{greenmar}{rgb}{0,.5,0}

\graphicspath{{./figures/}}

\title{Towards two-loop automation in OpenLoops}

\author[a]{Stefano Pozzorini}
\author[b]{Natalie Sch\"ar}
\author*[b]{Max F. Zoller}

\affiliation[a]{Universit\"at Z\"urich, CH-8057 Z\"urich, Switzerland}

\affiliation[b]{Paul Scherrer Institut, CH-5232 Villigen PSI, Switzerland}

\emailAdd{max.zoller@psi.ch}

\abstract{NLO scattering amplitudes are provided by fully automated numerical tools, such as OpenLoops, for a very wide range of processes. In order to match the numerical precision of current and future collider experiments, the higher precision of NNLO calculations is essential, and their automation in a similar tool a highly desirable goal.

In our approach, D-dimensional two-loop amplitudes are decomposed into Feynman integrals with four-dimensional numerators and (D-4)-dimensional remainders. The latter are reconstructed through process-independent rational counterterm insertions into lower-loop diagrams, while the first are expressed as loop momentum tensor integrals contracted with tensor coefficients. 

In this article, we describe a completely generic algorithm, first presented in \cite{Pozzorini:2022ohr}, for the efficient and numerically stable construction of these tensor coefficients. This algorithm is fully implemented in the OpenLoops framework for QED and QCD corrections to the Standard Model. For this implementation we present performance studies on numerical stability and CPU efficiency.}

\FullConference{%
  Loops and Legs in Quantum Field Theory - LL2022,\\
  25-30 April, 2022\\
  Ettal, Germany
}

\begin{document}
\maketitle

\section{Introduction}
\label{sec:intro}
Scattering amplitudes computed in perturbation theory are a key ingredient for Monte Carlo simulations of scattering processes at colliders.
Tree and one-loop amplitudes, required for LO and NLO predictions, can be obtained by numerical tools, such as 
\OpenLoops{} \cite{Cascioli:2011va,Buccioni:2017yxi,Buccioni:2019sur}, in a fully automated way.
In order to meet the precision requirements of the LHC and future colliders NNLO predictions, involving two-loop amplitudes, are essential for a wide range of processes. While dedicated NNLO calculations exist for many $2 \to 2$ and a few $2 \to 3$ processes, an automated NNLO tool would strongly augment the scope of NNLO phenomenology. 

 Scattering amplitudes at a given loop order $L$ are computed from Feynman diagrams $\Gamma$, 
\be
\calM_{L}(\heli) \,=\, \sum\limits_\Gamma \fullamp{L}{\Gamma}(\heli){},
\ee
where $\heli$ denotes the helicity configuration of the external particles of the process at hand. 
For processes with $\calM_{0}\neq 0$ the helicity and colour-summed squared tree-level amplitude
\be
\calW_{\ssst{LO}}\,=\,\f{1}{N_{\rm{hcs}}}
\sum\limits_{\heli,\col}
|\calM_{0}(\heli)|^2{},
\label{M2Wtree}
\ee
constitutes the LO contribution of the scattering probability density. 
Here, $1/N_{\rm{hcs}}$ encodes the average over initial-state helicity and colour d.o.f as well as symmetry factors
(see \cite{Buccioni:2019sur}).
A NLO calculation consists of a real-emission contribution, which has the same form as \eqref{M2Wtree} with one extra unresolved particle, and the virtual contribution
computed from the Born-loop interference
\be
\calW_{\ssst{NLO}}^{\ssst{virtual}}\,=\,\f{1}{N_{\rm{hcs}}}
\sum\limits_{\heli,\col} 
2\,\re \Big[\calM_{0}^*(\heli)\mathbf{R}\barM_{1}(\heli)\Big],
\label{M2Wone}
\ee
where $\mathbf{R}$ denotes the renormalisation procedure and the bar an amplitude computed in $D$ dimensions.

A NNLO calculation consists of three contributions. The double-real part has the same form
as \eqref{M2Wtree} with two extra unresolved particles, the 
real-virtual part has the same form as \eqref{M2Wone} with one extra particle, and the double-virtual contribution is given by
\be
\calW_{\ssst{NNLO}}^{\ssst{virtual}}\,=\, 
\sum\limits_{\heli,\col} 
2\,\re \Big[\calM_{0}^*(\heli)\,\mathbf{R}\barM_{2}(\heli)\Big] + |\mathbf{R}\barM_{1}(\heli)|^2.
\label{M2Wtwobar}
\ee

In the following, we will review the existing tree and one-loop \OpenLoops{} program, which provides all these contributions except for the
first term in \eqref{M2Wtwobar}. We will then
discuss the requirements for an automated two-loop tool, and  present a major building block of such a tool, namely a new algorithm for the numerical construction of two-loop integrands. 

\section{Tree-level and one-loop amplitude construction in OpenLoops}
\label{sec:oneloop}

In \OpenLoops{}, tree-level diagrams are constructed from subtrees $w_a$ through recursion steps \footnote{Subtrees are represented as blue bubbles in our graphs.}
\be
w^{\alpha}_a \,=\,
\parbox{0.13\textwidth}{\includegraphics[height=10mm]{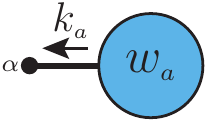}} \,=\,
\parbox{0.13\textwidth}{\includegraphics[height=20mm]{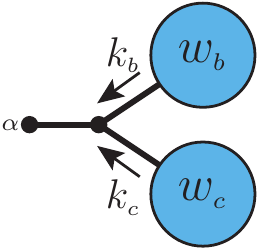}} \,=\,
\f{X_{\beta\gamma}^{\alpha}(k_b,k_c)}{\momk{a}^2-\mass{a}^2}
w^{\beta}_b 
w^{\gamma}_c ,
\ee
from two smaller subtrees, starting from the wave functions of the external particles. The kernel $X$ is derived from the Feynman rule of the connecting vertex and adjacent propagator with mass $\mass{a}$ and momentum $\momk{a}$. The recursion ends in connecting two subtrees into the full diagram. A high level of efficiency is achieved by recycling subtrees in multiple tree and loop diagrams.

Numerical tools construct the numerators of Feynman integrals in integer dimensions.
Typically, one-loop amplitudes $\barM_{1}$ in $D$ dimensions are split into an amplitude $\calM_{1}$ constructed from Feynman integrals with four-dimensional numerators and 
a remainder stemming from $(D-4)$-dimensional numerator parts.
The latter can be reconstructed through a finite set of process-independent rational counterterms \cite{Ossola:2008xq,
Draggiotis:2009yb,
Garzelli:2009is,
Pittau:2011qp}, which together with the one-loop UV counterterms are inserted into all possible tree-level diagrams, resulting in the counterterm amplitude $\calMCT_{0,1}(\heli)$. The renormalised $D$-dimensional amplitude is then computed as
\be
\mathbf{R}\barM_{1}(\heli) =
\calM_{1}(\heli)
+ \calMCT_{0,1}(\heli).
\ee

The amplitude of a one-loop diagram $\Gamma$ is given by
\be\calM_{1,\Gamma}(\heli) = \vcenter{\hbox{\scalebox{1.}{\includegraphics[height=27mm]{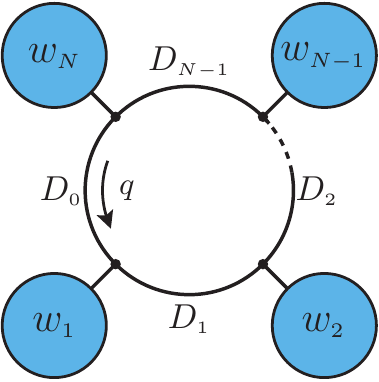}}}} 
= \colfac{1}{\Gamma}
\,\int\!\rd\momq\; \f{{\rm Tr}\left[\seg_1(q,\helisegment{1}{1})\!\cdots\!\seg_N({q,\helisegment{1}{N}})\right]}{ 
D_{0}\!\cdots\! D_{N-1}} \label{eq:oneloopdia}
\ee
with the colour factor\footnote{
Feynman diagrams with quartic vertices are split into colour-factorised contributions, each of which is treated as a separate diagram. This procedure is also applied at two loops. For details on the colour treatment in \OpenLoops{} see \cite{Buccioni:2019sur}.
} $\colfac{1}{\Gamma}$, the integration measure in loop momentum space 
$\int\!\rd\momq = \mu^{2\eps} \int \f{\rd^{^D}\! \bar
q}{(2\pi)^{^D}}$ and 
scalar propagator denominators $D_{a}(q)=(q+p_a)^2-m_a^2$
with mass $m_a$ and external momentum $p_a$.
The numerator factorises into loop segments with at most linear $q$-dependence,
\bea
\seg_a(q,\helisegment{1}{a})&=&\parbox{0.13\textwidth}{
\includegraphics[height=18mm]{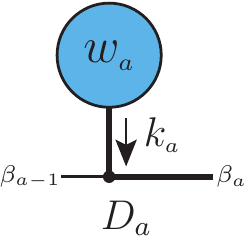}}
=
\{Y_{\sigma}^{a}+ Z_{\nu;\sigma}^{a}\,{q^\nu }
\}\, w^{\sigma}_a(\helisegment{1}{a}) ,
\eea
which consist of a loop vertex and propagator encoded in the universal building blocks $Y, Z$ and one or two external sub-trees $w_a$ with external momentum $k_a$ and helicity configuration $\helisegment{1}{a}$. Each segment is a matrix with Lorentz or spinor indices $\beta_{a-1}, \beta_a$, and the trace in \eqref{eq:oneloopdia} connects the indices
$\beta_{0}$ and $\beta_{N}$. The contribution of the diagram $\Gamma$ to \eqref{M2Wone} can be written as
\bea
\calW_{01,\Gamma} &=& \bar{\sum\limits_{\heli,\col}} 2\,\re \Big[\calM_{0}^*\calM_{1,\Gamma}\Big] 
= \re\left[
\int\!\rd\momq\;\f{\rm{Tr}\left[\calU_N(q)\right]}{ 
D_{0}\cdots D_{N-1}}\right]
\eea
where the integrand numerator 
\bea
\calU_N(q)
&=&
\sum\limits_{\heli}\calU_0(\heli)\prod\limits_{a=1}^N
\seg_a(q,\helisegment{1}{a})
\,=\,
\sum\limits_{\helisegment{1}{N}}
\cdots
\lb
\sum\limits_{\helisegment{1}{1}}\calU_0(\heli)\seg_1(q,\helisegment{1}{1})
\rb
\!\cdots\!
\seg_N({q,\helisegment{1}{N}})
\label{eq:UNdef}
\eea
factorises into nested helicity sums of segments and the Born--colour interference term
\bea
\calU_0(\heli) = 2\left(\sum\limits_{\col}\calM_0^*(\heli)\colfac{1}{\Gamma}\right).
\label{eq:Borncolour}
\eea

In \OpenLoops{}, a one-loop diagram is cut-opened at a chosen loop propagator $D_0$ and the resulting chain of segments in
\eqref{eq:UNdef}
is constructed recursively through steps\footnote{
The label $\helicheck_{k}$ denotes the helicity configuration of the external particles in the not yet attached segments $\seg_{k+1},\ldots,\seg_{N}$.}
\be
\calU_k(q,\helicheck_{k})
\,=\,
\sum\limits_{\helisegment{1}{k}}
\calU_{k-1}(q,\helicheck_{k-1})\seg_{k}(q,\helisegment{1}{k})
\ee
starting from $\calU_0(\heli)$, and employing an
on-the-fly summation of the segment helicities
$\helisegment{1}{k}$, introduced in \cite{Buccioni:2017yxi}. The numerator is systematically decomposed as
\be
\calU_k(q,\helicheck_{k})= \sum\limits_{r=0}^k
\calU_{k,\mu_1\ldots\mu_{r}}(\helicheck_{k})\,{q^{\mu_1}\ldots q^{\mu_{r}} },
\ee
and the numerical recursion is implemented at
the level of the tensor coefficients $\calU_{k,\mu_1\ldots\mu_{r}}$, retaining the analytical structure in $q$ throughout the amplitude construction. The tensor integrals in the resulting Born--loop interference
\be
\calW_{01,\Gamma}
 = 
\sum\limits_{r=0}^N
\calU_{N,\mu_1\ldots\mu_{r}}
\int\!\rd^D\!\!\momq\; \f{q^{\mu_1}\ldots q^{\mu_{r}}}{ 
D_{0}\!\cdots\! D_{N-1}}
\ee
are either reduced a posteriori, using external 
libraries such as Collier \cite{Denner:2016kdg}, or using the on-the-fly reduction method
\cite{Buccioni:2017yxi} with Collier or OneLoop \cite{vanHameren:2010cp} for the final evaluation of scalar integrals. This completely generic algorithm is fully implemented for QCD and EW corrections to the SM and available in the public \OpenLoops{} tool \cite{Buccioni:2019sur}.

\section{Two-loop amplitude construction}

Following the same strategy, we decompose the numerators of two-loop integrands into a part that can be numerically constructed in four dimensions, and $(D-4)$-dimensional remainders.
In
\cite{Pozzorini:2020hkx,Lang:2020nnl,Lang:2021hnw} it was demonstrated that the renormalised $D$-dimensional two-loop amplitude can be split into amplitudes and counterterms computed with four-dimensional loop numerators,
\be
\mathbf{R}\barM_{2}(\heli) = \calM_{2}(\heli) + 
\calMCT_{1,1}(\heli)+
\calMCT_{0,2}(\heli)+
\calMCT_{0,1,1}(\heli)
\ee
where the four terms on the rhs are the unrenormalised two-loop amplitude, 
the one-loop amplitude with one-loop rational and UV counterterm insertions,
the tree-level amplitude with two-loop rational and UV counterterm insertions, and the tree-level amplitude with double one-loop rational and UV counterterm insertions. The full set two-loop rational terms of UV origin were computed in \cite{Pozzorini:2020hkx,Lang:2020nnl,Lang:2021hnw} for QED and QCD corrections to the SM in a generic renormalisation scheme. Rational terms stemming from IR divergences are currently under investigation.

The calculation of the two-loop amplitude $\calM_{2}$ interfered with the Born $\calM_{0}$, is then split into the construction of the integrand at the level of loop momentum tensor coefficients, and the subsequent reduction and evaluation of the tensor integrals. In the following we present an algorithm for the first of these steps, focusing on the case of irreducible diagrams, which become 1PI on amputation of all external subtrees.\footnote{For reducible diagrams, which factorise into one-loop contributions, we refer to \cite{Pozzorini:2022ohr}, where a new and fully implemented algorithm based on the existing one-loop machinery is described.}

The amplitude of an irreducible two-loop diagram $\Gamma$ is given by 
\bea \calM_{2,\Gamma}(\heli)&=&
\vcenter{\hbox{\scalebox{1.}{\includegraphics[width=0.38\textwidth]{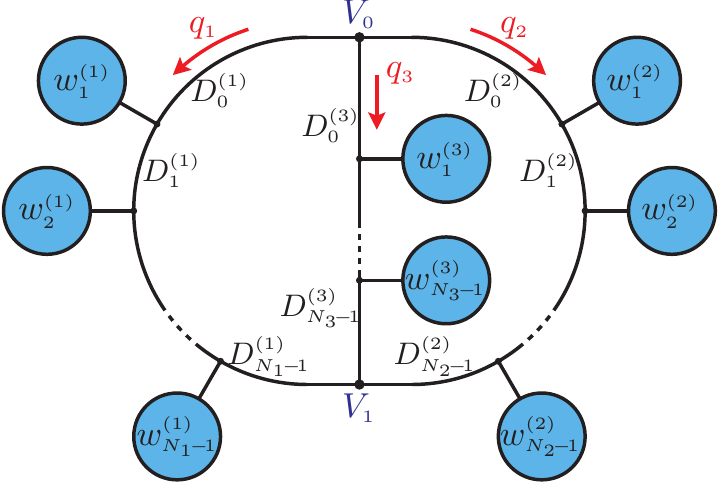}}}} 
=\colfac{2}{\Gamma}\!
\int\!\!\rd\momq_1\!\!\int\!\rd\momq_2
\f{\calN(q_1,q_2,\heli)
   }{\prod\limits_{i=1}^3\denc{i}} \Big|_{q_3=-(q_1+q_2)}
\eea
with the numerator of the integrand factorising into three chains connected by two vertices\footnote{For a quartic vertex $\vertex{j}$ the attached external subtree with
helicity $\helisegment{V}{j}$ is absorbed into the definition of $\vertex{j}$.}
$\vertex{0}, \vertex{1}$,
\bea
\calN({ q_1},{ q_2},\heli)= \prod\limits_{i=1}^{3}
\numc{i}({ q_i},\helic{i})
\prod\limits_{j=0}^{1}
   \vertex{j}({ q_1},{ q_2},\helisegment{V}{j}\,),
\label{eq:A2d}
\eea
and the three denominator chains $(i=1,2,3)$
factorising into scalar propagator denominators,
\bea
\label{eq:dendef}
\denc{i}&=&
D^{(i)}_0(\bar q_i)\cdots
D^{(i)}_{N_i-1}(\bar q_i)\,,
\qquad\mbox{where}\quad
D^{(i)}_a(\bar q_i) \,=\, \left(\bar q_i + p_{ia}\right)^2-m_{ia}^2\,.
\eea
Each chain depends on a single loop momentum and further factorises into segments of the same structure as at one loop,
\be
\numc{i}(q_i,\helic{i}) = \segment{i}{0}({q_i,\helisegment{i}{0}})\cdots\segment{i}{N_i-1}({q_i},\helisegment{i}{N_i-1}).
\ee
The helicity labels are defined in an additive way (see \cite{Pozzorini:2022ohr}), such that
\be 
\helig = \sum\limits_{i=1}^{3} \helic{i}
+\sum\limits_{j=0}^{1}\helisegment{V}{j},\qquad \helic{i}=\sum\limits_{a=1}^{N_i-1} \helisegment{i}{a}.
\ee
To compute the contribution of this diagram to \eqref{M2Wtwobar}, its colour factor
$\colfac{2}{\Gamma}$ is interfered with the Born,
\be
\calU_0(h)
 = 2\,\sum_{\col}\calM^*_0(\helig)\,\colfac{2}{\Gamma},
\label{eq:colborn_two}
\ee
to form a building block in the construction of the numerator of the Born two-loop interference, 
\be
\calU({q_1},{q_2}) = \sum\limits_{h} \calU_0({h})\,\calN({q_1},{q_2},{h})\label{eq:num_interf}
=
\sum_{r_1=0}^{R_1}\sum_{r_2=0}^{R_2} 
{\calU}_{\mu_1 \cdots \mu_{r_1} \nu_1 \cdots \nu_{r_2}}
q_1^{\mu_1}\cdots q_1^{\mu_{r_1}}\,q_2^{\nu_1}\cdots q_2^{\nu_{r_2}}.
\ee
This construction is again performed at the level of the loop momentum tensor coefficients.
The most efficient recursive algorithm was found through a cost analysis of possible candidates
for a wide range of processes \cite{Pozzorini:2022ohr}. This completely generic algorithm is fully implemented and validated for QED and QCD corrections to SM processes. It consists of the following steps:
\paragraph{0.} The chains are sorted by length, such that $N_1 \geq N_2 \geq N_3$. The order of $\vertex{0}$ and $\vertex{1}$ is determined by vertex type, in a way that the efficiency of the subsequent steps is maximised (see \cite{Pozzorini:2022ohr} for details).
\paragraph{1.} The shortest chain is constructed through the recursion ($n=0,\ldots,N_3-1$)
 \be
 \numpc{3}{n}({ q_3},{ \helipc{3}{n}}) = \numpc{3}{n-1}({ q_3},{ \helipc{3}{n-1}})\cdot\segment{3}{n}({ q_3},{ \helisegment{3}{n}}) 
 \quad\text{with} \quad
 \helipc{3}{n} = \sum\limits_{a=1}^{n} \helisegment{3}{a},
 \ee
 recycling intermediate results in multiple Feynman diagrams.
\paragraph{2.}
 The full diagram interfered with the Born is then constructed through a sequence of recursions starting from 
 $\numpi{-1}(\helig)=\calU_0(\helig)$ defined in \eqref{eq:colborn_two}:
 \begin{itemize}
    \item[2.1] The longest chain, which usually contains the majority of helicity d.o.f., is constructed as
\be    
\numpi{n}({ q_1},{ \helipcc{1}{n}})
=
{\sum\limits_{\helisegment{1}{n}}}
\numpi{n-1}({ q_1},{ \helipcc{1}{n-1}})\cdot\segment{1}{n}({ q_1},{ \helisegment{1}{n}})
\quad 
\text{with}\quad
{ \helipcc{1}{n}= \helig -\sum\limits_{a=1}^n\helisegment{1}{a}}
\ee
and $n=0,\ldots,N_1-1$. The on-the-fly summation of a large fraction of the helicity configurations at a stage where the intermediate results depend on a single loop momentum $q_1$ is an important reason for the efficiency of the algorithm.
\item[2.2] The vertex $\vertex{1}$ is connected,
\be
\numinter{1}({ q_1},{ q_3},{ \helic{2}+\helisegment{V}{0}}) 
=
{ \sum\limits_{\helic{3},\helisegment{V}{1}}}\;
\numpi{N_1-1}({ q_1},{ \helig-\helic{1}})\;
\numpc{3}{N_3-1}({ q_3},{ \helic{3}})\;
\vertex{1}({ q_1},{ q_3},\helisegment{V}{1}),
\ee
summing the helicities of the previously constructed $\numc{3}$ and of $\vertex{1}$. The introduction of a second loop momentum strongly increases the complexity of this and the following steps, which is counterbalanced to a large degree by the on-the-fly helicity summation.
\item[2.3] The vertex $\vertex{0}$ is connected,
\be
\numpd{-1}({ q_1},{ q_2},{ \helic{2}})
= 
{ \sum\limits_{\helisegment{V}{0}}}\;
\numinter{-1}({ q_1},{ q_3},{ \helic{2}+\helisegment{V}{0}})\;
\vertex{0}({ q_1},{ q_2},\helisegment{V}{0})\Big|_{ q_3= -(q_1+q_2)},
\ee
reducing the number of open Lorentz/spinor indices from three to two.
\item[2.4] The last chain is constructed through the recursion $(n=0,\ldots,N_2-1)$
\be
\numpd{n}({ q_1},{ q_2},{ \helipccloc{2}{n}}) 
=
{ \sum\limits_{\helisegment{2}{n}}}
  \numpd{n-1}({ q_1},{ q_2},{ \helipccloc{2}{n-1}}) \; \segment{2}{n}({ q_2},{ \helisegment{2}{n}})
  \quad 
\text{with}\quad
  \helipccloc{2}{n}=\sum\limits_{a=n+1}^{N_2-1} \helisegment{2}{a},
\ee
resulting in 
$\calU({q_1},{q_2}) = \numpd{N_2-1}({ q_1},{ q_2},0)$, where all helicities are summed.
 \end{itemize}

\section{CPU efficiency and numerical stability} \label{sec:CPUeff}

\begin{figure}[t]
\centering
\includegraphics[trim = 5mm 1mm 5mm 14mm, clip, height=78.5mm]{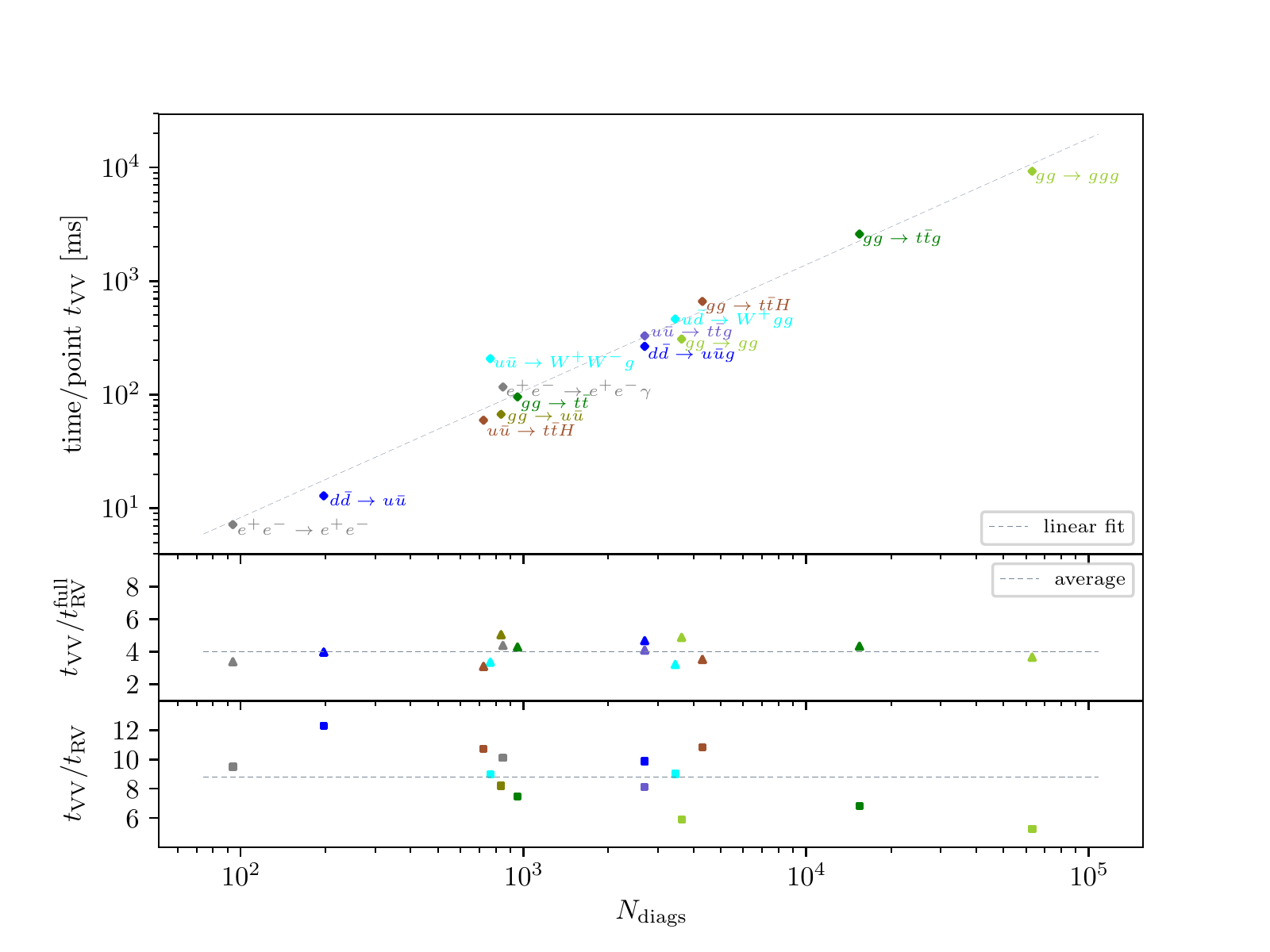}
\caption{Computation time for selected SM processes with massless fermions. For processes with external $e^{\pm}$, QED corrections were considered, for all other processes QCD corrections. Fermions in loops are considered massless, except for the top which is included in processes with an external top.}
\label{fig:speed}
\end{figure}

The CPU efficiency of this new algorithm was measured for a wide range of QED and QCD 
processes.\footnote{For each process the time for the construction of all tensor coefficients of the irreducible diagrams for $1000$ uniform random points was measured on a machine with a single Intel i7-6600U @ 2.6 GHz processor and 16GB RAM.} The average time $t_{\ssst{VV}}$ per phase space point against the number of Feynman diagrams $N_{\ssst{diags}}$ is shown in the upper plot of Fig.~\ref{fig:speed}. The runtimes  range from a few ms for simple QED and QCD processes to $9.2$s for $gg\to ggg$. The computation time scales linearly with the number of diagrams.
In order to estimate the CPU cost of this algorithm in relation to a full NNLO calculation, we compare it to the construction time for the tensor coefficients of the real--virtual contribution, $t_{\ssst{RV}}$, and for the full real-virtual probability density, $t_{\ssst{RV}}^{\ssst{full}}$, including the one-loop tensor integrals. The 
lower plots of Fig.~\ref{fig:speed}, show approximately constant ratios for the considered processes with 
\mbox{$t_{\ssst{VV}}/t_{\ssst{RV}} = 9 \pm 3$} and \mbox{$t_{\ssst{VV}}/t_{\ssst{RV}}^{\ssst{full}} = 4 \pm 1$}.
We conclude that the cost of the
double-virtual and real-virtual tensor coefficient construction is comparable, in particular considering that more evaluations of the real-virtual part are usually required during a Monte Carlo integration over the phase space.

The implementation of the algorithm also shows high numerical stability, as demonstrated by relative uncertainty measurements for $2\to 2$ and $2\to 3$ QCD amplitudes computed in double precision for $10^5$ uniform random points. These relative uncertainties are in the range of $10^{-16}$ to $10^{-14}$ for the bulk of the points, and never below order $10^{-12}$ and $10^{-11}$ (for details see \cite{Pozzorini:2022ohr}).
\section{Conclusion}
We presented a new and completely generic algorithm for the CPU efficient and numerically stable construction of the loop-momentum tensor coefficients of two-loop amplitudes. This 
constitutes a key building block in the development of an automated two-loop \OpenLoops{} tool.
\section*{Acknowledgements}
This research was supported by the Swiss National Science Foundation (SNSF) under the Ambizione grant PZ00P2-179877. 
The work of S.P. was supported through contract BSCGI0-157722.


\end{document}